\documentclass[twocolumn,showpacs,preprintnumbers,amsmath,amssymb]{revtex4}
\usepackage[dvips]{graphicx}

\begin{document}

\title{ Neutron Skin Thickness of ${}^{90}$Zr 
Determined By  Charge Exchange Reactions}   
\author{ K. Yako~$^{\rm a}$,
%\footnote{Email:~yakou@phys.s.u-tokyo.ac.jp
%Tel:~+81-3-5841-4236, Fax:~+81-3-5841-7642
%}
H. Sagawa~$^{\rm b}$, and H. Sakai~$^{\rm a}$}
\address{
${}^{\rm a}$ Department of Physics, University of Tokyo, Bunkyo, Tokyo 113-0033,~Japan\\
${}^{\rm b}$~Center for Mathematical Sciences, the University of Aizu\\
Aizu-Wakamatsu, Fukushima 965-8580, Japan
}
\begin{abstract}
Charge exchange spin-dipole (SD) excitations 
of ${}^{90}{\rm Zr}$ are studied by the ${}^{90}{\rm Zr}(p,n)$ and 
${}^{90}{\rm Zr}(n,p)$ reactions at 300 MeV.
A multipole decomposition technique is employed to obtain
the SD strength distributions in the cross section spectra.
For the first time, a model-independent 
SD sum rule value is obtained: $148\pm12~{\rm fm^2}$.
The neutron skin thickness of ${}^{90}{\rm Zr}$ is determined to be 
$0.07\pm0.04~{\rm fm}$ from the SD sum rule value.
%\medskip

%\centerline{ Keywords: Spin-dipole state} 
\end{abstract}
\maketitle

\newpage
        Proton and neutron distributions are
among the most fundamental properties of nuclei.
        Proton distributions are precisely known
 from the charge distributions determined by electron scattering~\cite{Frois}.
        On the other hand, our knowledge of neutron
 distributions, which have been studied mainly by hadron-nucleus
 scattering, is limited
 because  descriptions of strong interactions in nuclei are highly 
 model-dependent~\cite{pscatt2}.
        Reliable neutron distributions 
 will improve the understanding of the nucleus and nuclear 
 matter~\cite{Brown, Furnstahl, Prakash, Steiner, Dani}.
        Recent theoretical studies  using the Skyrme Hartree-Fock (HF) 
 and relativistic mean-field models~\cite{Brown,Furnstahl,Yoshi04} 
 have shown that the neutron skin thickness,
 defined as the difference between the root mean square (rms) radii of the 
 proton and neutron distributions,
 imposes a strict constraint on
 the neutron matter equation of state,
 which is an important ingredient in studies of 
 neutron stars~\cite{Prakash, Steiner}.  
 It is also known that the neutron skin 
 thickness is strongly correlated with the nuclear 
 symmetry energy~\cite{Dani,Yoshi06}.
        Reliable neutron distributions are also needed for 
 analyses of atomic parity violation experiments~\cite{apvexp1, Chen} 
 and of pionic states in nuclei~\cite{PionicAtom}.

   Several attempts have been
 made to determine neutron 
 distributions~\cite{pscatt1,pscatt2,pscatt3,GDR,Trzcinska}.
     Ray {\it et al.}\ analyzed proton elastic scattering on several nuclei
 at 800 MeV
  using impulse approximation and obtained a neutron thickness of
 $0.09\pm0.07$~{\rm fm} for ${}^{90}{\rm Zr}$~\cite{pscatt1}.
      The cross sections for excitation of isovector giant dipole resonances 
 with alpha scatterings were measured at KVI and 
 neutron skin thicknesses of ${}^{116,124}{\rm Sn}$ and ${}^{208}{\rm Pb}$ 
 were obtained~\cite{GDR} with uncertainties of $\pm 0.12~{\rm fm}$.
        Trzci\'{n}ska {\it et al.}\ measured 
the strong-interaction effects on 
antiprotonic x-rays on various targets ranging from ${}^{16}{\rm O}$ to 
${}^{238}{\rm U}$ and obtained a neutron skin thickness of $0.09~{\rm fm}$ for 
${}^{90}{\rm Zr}$ with a statistical error of $\pm0.02~{\rm fm}$~\cite{Trzcinska}. 
     Unfortunately, these analyses are model-dependent.
        Parity violation electron scattering is a 
 promising tool for observing the neutron distributions 
 cleanly~\cite{JLab,escatt}, although no data are available so far. 

        An alternative method for determining the 
 neutron rms radius is provided by 
 the model-independent sum rule strength of charge exchange 
 spin-dipole (SD) excitations~\cite{Gaarde}.
     The operators for SD transitions are defined by
\begin{eqnarray}
\hat{S}_{\pm} &=& \sum_{im\mu} t_{\pm}^{i}\sigma_{m}^{i} r_{i}Y_{1}^{\mu}
(\hat{r}_{i})
\label{eq:ope_sd}
\end{eqnarray} 
 with the isospin operators  $t_{3} =  t_{z}$, 
 $t_{\pm} =  t_{x}\pm it_{y}$. 
        The model-independent sum rule is derived as
\begin{equation}
 S_{-}- S_{+}
 = \frac{9}{4\pi}\left(N\langle r^2\rangle _n -Z\langle r^2\rangle _p\right),
 \label{eq:sum_sd}
\end{equation}
where $S_\pm$ are the total SD strengths.
The mean square radii of the neutron and proton distributions are 
denoted as $\langle r^2\rangle _n$ and $\langle r^2\rangle _p$,
respectively.
        Thus, the rms radius of the neutron distribution 
 $\sqrt{\langle r^2\rangle _n}$ 
or the neutron skin thickness $\delta_{np}=\sqrt{\langle r^2\rangle_n}
           -\sqrt{\langle r^2\rangle_p}$
 can be derived from 
 Eq.~(\ref{eq:sum_sd}) by using the rms radius of the proton distribution
 $\sqrt{\langle r^2\rangle _p}$ obtained from the charge radius
 if the sum rule value ($S_--S_+)$ is obtained experimentally.

        To obtain the SD sum rule value, Krasznahorkay {\it et al.}\ measured
the $({}^3{\rm He},t)$ reaction on tin isotopes at 450 MeV~\cite{SD-Pb}.
        The cross section spectra at $\theta = 1^\circ$ 
were analyzed by a peak fitting technique assuming the Lorentzian shape, 
and the $S_-$ value 
was obtained.
        Since there was no $(n,p)$-type measurement, they 
determined the $S_+$ value by 
assuming the energy-weighted sum rule in a simple model
where the unperturbed particle-hole ({\it ph})
energies are degenerate at a certain energy~\cite{Gaarde}.
        The overall normalization of the sum rule value was done 
by using the calculated result of neutron skin thickness of 
${}^{120}{\rm Sn}$.        
	To determine the sum rule value model-independently, however, 
one needs to perform $(n,p)$-type experiments as well as to identify the 
$\Delta L =1$  cross sections in the observed spectra.
  Then, one needs to relate the $\Delta L =1$ 
cross sections to the SD strengths.

        This paper represents the first attempt to extract the 
model-independent sum rule value from both $(p,n)$ and $(n,p)$ 
reactions on ${}^{90}{\rm Zr}$.
        The dipole components of the cross section spectra 
 are identified by multipole decomposition (MD) 
 analysis~\cite{review}
 of the ${}^{90}{\rm Zr}(p,n)$~\cite{Wakasa} and
 ${}^{90}{\rm Zr}(n,p)$~\cite{Yako} data at 300~MeV.
    It should be noted that 
 at 300 MeV, the spin-flip cross sections are large, while the 
 distortion effects are 
 minimal~\cite{Shen}.  
        Thus, the characteristic shapes of the angular 
 distributions for each angular momentum transfer ($\Delta L$) 
 are most distinct. 
    The contribution of the non-spin dipole component is small
 due to the energy dependence of the effective interaction.

     The MD analyses were 
 performed on the $(p,n)$ and $(n,p)$ excitation energy spectra 
 in a consistent manner~\cite{Yako, review} to obtain 
 various $\Delta L$ components of the cross section. 
    We use here the extracted $\Delta L=1$ cross section spectra 
 given in Fig.~2 of Ref.~\cite{Yako}.
    The SD strengths are obtained by assuming a proportionality relation
 similar to that established for Gamow-Teller excitation.
    The proportionality relation between $B({\rm SD})$ and 
 the $\Delta L=1$ component of the cross section, 
 $\sigma_{\Delta L=1,{\pm}}(q,\omega)$, is given by 
\begin{equation}
        \sigma_{\Delta L=1,{\pm}}(q,\omega)
        =\hat{\sigma}_{\rm SD_\pm}(q,\omega)B({\rm SD_\pm}),
\label{eq:proportionality}
\end{equation}  
where 
$\hat{\sigma}_{\rm SD_\pm}(q,\omega)$ is the SD cross section per 
$B({\rm SD_\pm})$ and depends on the momentum transfer $q$ and the 
energy transfer $\omega$.
        The $\sigma_{\Delta L=1}(q,\omega)$ data for the $(p,n)$ and 
$(n,p)$ channels were taken from the result of MD analysis at $4.6^\circ$ 
and 4--$5^\circ$, respectively.
        The $\sigma_{\Delta L=1}(q,\omega)$ spectra at these angles 
 are most sensitive to the SD cross sections since the corresponding 
 momentum transfers are $q=0.3$--$0.4~{\rm fm^{-1}}$ and thus,
 the $\Delta L=1$ cross sections take on maximum values.

      Distorted wave impulse approximation (DWIA) calculations are
used to obtain $\hat{\sigma}_{\rm SD_\pm}(q,\omega)$.
The DWIA calculations are performed with the computer code 
DW81~\cite{dw81} for the $0^-$, $1^-$, and $2^-$ transitions.
The one-body transition densities are calculated from 
pure 1{\it p}1{\it h}
configurations.
        All $1\hbar\omega$ configurations were  examined.
        The optical model potential (OMP) parameters 
 are taken from Ref.~\cite{CooperHama}.
        The effective {\it NN} interaction is taken from the $t$-matrix
parameterization of the free {\it NN} interaction by Franey and Love at
 325 MeV~\cite{FL}.
        The radial wave functions are generated from a Woods-Saxon 
 potential~\cite{BM}, adjusting the depth of the central potential
 to reproduce the binding energies.
 Details of the calculations are found in Refs.~\cite{review, Yako}.

        The calculated ${\rm SD}_-$ unit cross sections 
 at $\theta=4.6^\circ$ and $\omega = 0$~MeV are 
 $0.28\pm0.03$, $0.24\pm0.06$, and 
 $0.29\pm0.05~{\rm mb/sr/fm^2}$ for the 
 $0^-$, $1^-$, and $2^-$ transitions, respectively.  
        The uncertainty indicated for each multipole is
 the dependence of the cross sections on the {\it ph} configurations.
        Gaarde {\it et al.}\ pointed out that the tensor term of the
effective {\it NN} interaction affects the proportionality 
among the multipoles in the study of 
SD excitations in the ${}^{12}{\rm C}(p,n)$ reaction at 160 MeV.
        At 300 MeV, however, 
 the ratio of the amplitude of 
 the isovector tensor interaction to that of the isovector spin 
 $(\sigma\tau)$ interaction at $q=0.3$--$0.4~{\rm fm^{-1}}$ is 
 smaller~\cite{FL} so that 
 the calculated unit cross sections are 
 close to one another.
        To check the validity of the above calculations
 with pure 1{\it p}1{\it h} configurations,
 the calculations of SD unit cross sections are 
 performed by using the transition densities for several states obtained 
 by the HF+RPA (random phase approximation) calculations~\cite{SY06}. 
        The calculated unit cross sections are found to be consistent with
 the above values.

        Since the main subject of this study is
 the total SD strength, rather than the 
 individual strength of each transition,
 we use the averaged unit cross section of 
 $\hat{\sigma}_{\rm SD_-}(4.6^\circ,0~{\rm MeV})=0.27~{\rm mb/sr/fm^2}$ 
 in the analysis.
        Similarly, the unit cross section in the ${\rm SD}_+$ channel 
 is estimated to be 
 $\hat{\sigma}_{\rm SD_+}(4{\rm -}5^\circ,0~{\rm MeV})=0.26~{\rm mb/sr/fm^2}$.
        The unit cross sections are calculated at each 
 energy bin of the cross section histogram.
        The systematic uncertainty in 
 $\hat{\sigma}_{\rm SD_\pm}(q,\omega)$ due to the input parameters for DWIA
 calculations has been evaluated by using  harmonic oscillator 
 radial wave functions and by other sets of OMP~\cite{Shen, DParis}. 
 The systematic uncertainty thus estimated is 14\%.
\begin{figure}[t]
\includegraphics[width=3.3in,clip]{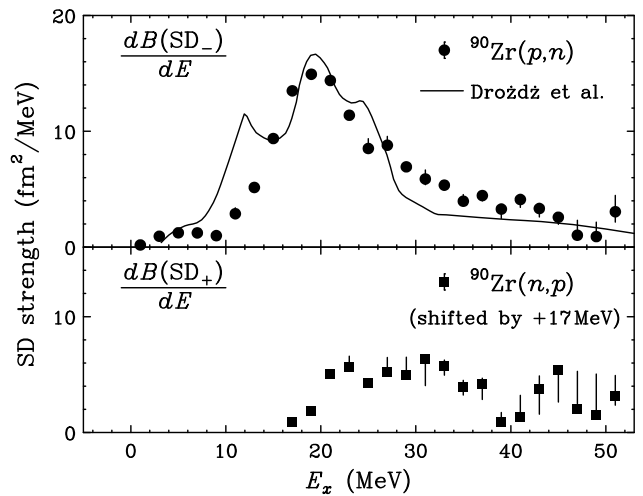}
\caption{\label{fig:zr90_sd}
Charge exchange SD strength $\frac{dB({\rm SD_-})}{dE}$ (upper panel) and 
$\frac{dB({\rm SD_+})}{dE}$ (lower panel).
The circles and squares are the experimental data.
The $\frac{dB({\rm SD_+})}{dE}$ spectra are shifted by +17~MeV.
The curve is the results of the second RPA calculation by 
Dro\.{z}d\.{z} {\it et al.}~\cite{Dro87}. 
}
\end{figure}

        The $\frac{dB({\rm SD})}{dE}$ distributions obtained by using 
 Eq.~(\ref{eq:proportionality}) are shown in Fig.~\ref{fig:zr90_sd}.
        The horizontal axis in the $\frac{dB({\rm SD_-})}{dE}$ spectrum 
 is the excitation energy of the residual ${}^{90}{\rm Nb}$ nucleus.
        The ${\rm SD_-}$ strength spectrum shows a dominant resonance structure
 centered at $E_x=20$~MeV and the strength extends to $\sim 50$~MeV excitation.
        The $\frac{dB({\rm SD_+})}{dE}$ distribution is shifted by $+17$ MeV to account 
 for the Coulomb displacement energy~\cite{BM} and the 
 nuclear mass difference.
        The ${\rm SD_+}$ strength distribution forms a broad bump 
 centered at 28~MeV (or 11~MeV in ${\rm {}^{90}Y}$) with a width
 of $\sim15$~MeV.

        Dro\.{z}d\.{z} {\it et al.}\ studied the ${\rm SD_-}$ strengths 
with the RPA model including the coupling between 1{\it p}1{\it h} and 
2{\it p}2{\it h} states,
though the calculation is not self-consistent, using a phenomenological 
Woods-Saxon potential to obtain the ground state wave function~\cite{Dro87}.
        They found that the mixing of 2{\it p}2{\it h} states 
 results in a large asymmetric spread 
 in the strength of the SD resonances, with about 30$\%$ of the 
 total strength shifted to excitation energies above 28 MeV~\cite{Dro87}.
        The curve in Fig.~1 shows the predicted strength distribution.
        This calculation gives a reasonable description of the 
 strengths above 20 MeV although it overestimates the strengths at lower 
 excitation energies below 15~MeV.
	 To discuss the details of SD strength distributions 
 and their relations to the neutron skin thickness, 
 the self-consistent HF+RPA models should be employed where 
 the same two-body interaction is used throughout the calculations.

\begin{figure}[t]
\includegraphics[width=3.3in,clip]{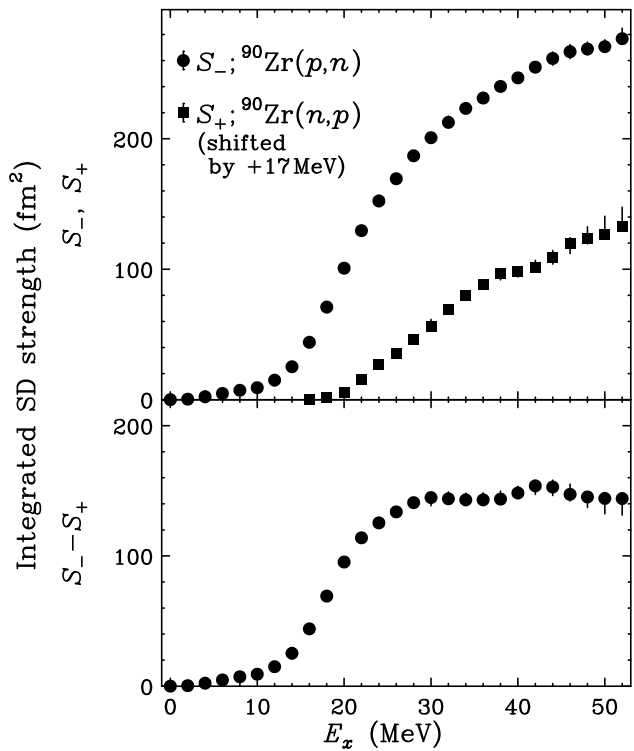}
\caption{\label{fig:zr90_sd_sum}
Integrated charge exchange SD strengths.
The upper panel shows the $S_-$ and $S_+$ spectra.
The lower panel shows the $S_--S_+$ spectrum.
}
 \end{figure}
The integrated SD strength, 
\begin{equation}
S_\pm \equiv \int_0^{E_x}\frac{dB({\rm SD_\pm})}{dE}dE,
\end{equation}
 is plotted 
in Fig.~\ref{fig:zr90_sd_sum}.
  While both integrated SD strengths increase steadily,
the sum rule value, $S_--S_+$, remains almost constant in the excitation 
energy range of 30--50~MeV, where it lies within 
$149\pm5~{\rm fm}^2$.
  Since the MD analysis in the $(n,p)$ channel is unstable above an
 excitation energy of 40 MeV in Fig.~\ref{fig:zr90_sd},
 we integrate the strengths up to 40 MeV.
  The ${\rm SD}_-$ strengths are $S_- = 247\pm4({\rm stat.})\pm12({\rm MD})~{\rm fm^2}$,
where the statistical uncertainty and the uncertainty in the MD analysis 
are 
given.
        The corresponding $S_+$ value is 
$98\pm4({\rm stat.})\pm5({\rm MD})~{\rm fm^2}$, integrated to an
 excitation energy of 23~MeV in $^{90}$Y (40~MeV in Fig.~\ref{fig:zr90_sd_sum}).
        The sum rule value yields
 $S_--S_+=148\pm6({\rm stat.})\pm7({\rm syst.})\pm7({\rm MD})~{\rm fm^2}$,
         where the systematic uncertainty of the normalization in the 
 cross section data (5\%) is also included.
         The 14\% uncertainty in the SD unit cross section is not included.

     If this sum rule value is interpreted in terms of Eq.~(\ref{eq:sum_sd}), 
we obtain
$N\langle r^2\rangle_n -Z\langle r^2\rangle_p = 207\pm17~{\rm fm^2}$,
 where the statistical and systematic uncertainties 
 are combined in quadrature
 with
 the uncertainty in
 MD analysis.  
     The rms radius of the proton matter in ${}^{90}{\rm Zr}$ is estimated
 to be 4.19~fm after correcting for the effect of the proton form factor
 from the charge radius~\cite{Vries}.
     The neutron skin thickness and the rms radius of the neutron 
 distribution calculated from Eq.~(\ref{eq:sum_sd})
 are $\delta_{\it np}=0.07\pm0.04~{\rm fm}$ and 
 $\sqrt{\langle r^2\rangle_n}=4.26\pm0.04~{\rm fm}$, respectively.
     The $\delta_{np}$ value obtained in the present study is
 consistent with that
 obtained from the analysis of  proton elastic scattering
 ($0.09\pm0.07~{\rm fm}$)~\cite{pscatt1} 
 but with a smaller uncertainty.
        Our results also agree with the neutron skin thickness determined by
 antiprotonic x-ray measurements by Trzci\'{n}ska {\it et al.} 
 ($0.09\pm0.02~{\rm fm}$)~\cite{Trzcinska},
 though their analysis contains some assumptions whose uncertainties
 are not specified.
     We note that the accuracy of 
 $\sqrt{\langle r^2\rangle_n}$ in 
 this study is 1\% level, which is the same as the goal of the 
 parity violation experiment of electron scattering 
 at Jefferson Laboratory~\cite{JLab}.
     To improve the reliability of the present analysis, 
 the SD unit cross sections should be studied in the mass region 
 around 90 by measuring the cross sections of the SD transitions
 whose strengths are known from decay measurements.

        In summary, 
        a consistent MD analysis of the $(p,n)$ 
and $(n,p)$ reaction data from ${}^{90}{\rm Zr}$ has been performed 
and SD strength distributions of both channels are obtained experimentally
  for the first time. 
        The integrated strengths obtained experimentally are
$S_- = 247\pm4({\rm stat.})\pm4({\rm syst.})\pm12({\rm MD})~{\rm fm^2}$ and 
$S_+ = 98\pm4({\rm stat.})\pm4({\rm syst.})\pm5({\rm MD})~{\rm fm^2}$, up to
 40 MeV and 23 MeV, respectively. 
        By using the two experimental 
 sum rule values, the model-independent formula (\ref{eq:sum_sd}) 
 yields a $N\langle r^2\rangle_n -Z\langle r^2\rangle_p$ value of 
 $207\pm17~{\rm fm^2}$,  
 which corresponds to a neutron skin thickness of $\delta_{np}$=
 $0.07\pm0.04~{\rm fm}$.
        The method used in this work is applicable to heavy and medium heavy
 nuclei on which both the $(p,n)$ and the $(n,p)$ measurements are possible.

        The authors would like to thank all members of the 
RCNP-E149 collaboration.
        The authors are grateful to Prof. I.~Hamamoto, Prof. G.~Col\`{o}, 
and Prof. T.~Wakasa for useful discussions. 
        This work was supported in part by Grants-in-Aid for Scientific 
Research Nos.\ 10304018, 16540259, and 17002003 of 
the Ministry of Education, Culture, Sports, Science, 
and Technology of Japan.

\end{document}